\documentstyle[aps]{revtex}
\begin{document}
\draft
\tighten

\title{Finite temperature excitations of Bose gases in anisotropic traps}
\author{J\"urgen Reidl$^1$, Andr\'as Csord\'as$^2$, Robert Graham$^1$, 
P\'eter Sz\'epfalusy$^3$}
\address{
$^1${\it Fachbereich Physik, Universit\"at Gesamthochschule Essen, 
45117 Essen, Germany}\\
$^2${\it Research Group for Statistical Physics of the 
Hungarian Academy of Sciences,\\
P\'azm\'any P\'eter s\'et\'any 1/A, H-1117 Budapest, Hungary}\\
$^3${\it Department of Physics of Complex Systems, 
E\"otv\"os University, P\'azm\'any P\'eter s\'et\'any 1/A, 
H-1117 Budapest, Hungary,\\
and 
Research Institute for Solid State Physics and Optics, 
P.O. Box 49, H-1525 Budapest, Hungary}}

\date{\today}
\maketitle
\vspace{1cm}
\begin{abstract}
The mode frequencies of a 
weakly interacting Bose gas in a  magnetic trap are studied 
as a function of the anisotropy of the trap.
As in earlier works 
the generalized {\it Hartree-Fock-Bogoliubov}  equations
within the Popov approximation ({\it HFB-Popov}) 
are used for our calculations. 
The new feature of our work is the combined use of a mode expansion in a 
finite basis and a semiclassical approximation of the highly excited states.
The results are applied to check the accuracy of the  
recently suggested  {\it equivalent zero-temperature condensate} 
(EZC) approximation which involves a much simpler model.

\end{abstract}

\section{INTRODUCTION}

New experiments \cite{Jin1,Stamper,Mewes,Jin2} on condensed Bose gases 
with oscillating trap potential
permit the excitation   of low-lying collective modes with given symmetry. 
The measurement   of their 
excitation frequencies provides a good opportunity
to analyze the applicability of different approximations and numerical 
approaches in many-body-theory.
Zero temperature calculations 
with different number of atoms  in the condensate 
\cite{Edwards,You} are in 
excellent agreement with the corresponding spectra measured in 
the JILA-TOP  \cite{Jin1} and the MIT trap \cite{Mewes}.
The frequently used starting point for the theoretical studies 
are the coupled Bogoliubov equations. 
Their  eigenvalues can be  determined  numerically 
by expanding the solutions in a basis-set of 
orthonormal functions \cite{HutchZarGriffin,Isoshima,HutchZar}.
Analytical solutions of the Bogoliubov equations 
have been obtained in the hydrodynamic limit by {\it Stringari} 
\cite{Stringari} and in \cite{Oeb,Fliesser}.

For finite temperatures a possible and frequently used  
extension of the Bogoliubov 
equations are 
the Hartree-Fock-Bogoliubov equations
with neglection of the anomalous expectation values,
commonly  referred to as
Hartree-Fock-Bogoliubov (HFB)-Popov equations \cite{Popov,Griffin}.
Without their neglection 
the anomalous expectation values appearing in the self-consistent 
Hartree-Fock theory would lead to an unphysical  
energy gap of the collective modes in a spatially homogeneous 
Bose-condensed system violating the Hugenholtz-Pines theorem \cite{HohMa}.
In a spatially inhomogeneous trapped Bose-condensate the Hugenholtz-Pines 
theorem does not apply, but in this case the Kohn-theorem \cite{Kohn} would be 
violated, which states that the dipole excitations have the frequencies of the
harmonic trap.

The HFB-Popov equations have been solved within 
the local density approximation \cite{Giorgini,Dalfovo},
which is sufficient to determine the thermodynamic properties of the trapped
Bose-condensates.
Beyond that,
similar to the numerical method used in the case of 
the zero-temperature Bogoliubov equations,
the excitation spectra can be  derived  by an expansion in a basis set of 
eigenfunctions.
{\it Dodd et al.} \cite{DoddEd} present the numerical results together with 
the experimental data. 
For temperatures    $T\le 0.65\, T_0 $ ($T_0$ is the 
theoretical transition temperature for the ideal, trapped Bose gas) 
the calculated values differ by less than 
$5\%$ from the experiment.
On the other hand, it was also pointed out in \cite{DoddEd} that 
the same accuracy between numerics and experiment is already obtained  
by  a much simpler approximation.
The excitation spectra at finite temperatures  
are simply derived by a zero temperature calculation but with 
the number of atoms in the trap reduced to account for 
the thermal depletion.  
Correspondingly, this solution is denoted as 
{\it equivalent zero-temperature condensate} (EZC)-solution \cite{DoddEd}.

In the present paper we present a further numerical study of the 
HFB-Popov equations with two aims: On the one hand we wish to calculate the
excitation spectra by combining the use of a basis set expansion for the 
low-lying modes with the local density approximation for the high-lying modes
beyond a suitably chosen energy cut-off. This eliminates truncation errors
while keeping the discrete basis set of modes reasonably small, thereby 
speeding up our computation. The second aim then is to use this gain in 
efficiency for  a more detailed check of the accuracy of the EZC-approximation
in comparison with the HFB-Popov approximation.
In particular we shall provide a comparison for axially symmetric traps 
with 
arbitrary anisotropy 
both for low lying excitations and also for excitations at energies
above the chemical potential.

\section{THEORY}

The HFB-Popov mean-field theory for inhomogeneous dilute Bose gases has 
been derived in detail by Griffin in Ref. \cite{Griffin}. 
We merely give a brief summary of the basic equations.
The Bose gases in the experiment are 
weakly interacting and dilute. Therefore, 
the  s-wave approximation $v(\bbox{r}-\bbox{r}')=
g\,\delta(\bbox{r}-\bbox{r}')$ is adequate to 
describe the interaction where $g=4\pi\hbar^2a/m$ and $a$ is the s-wave
scattering length.
To the Bose field operator  $\hat{\psi}(\bbox{r},t)$
the usual decomposition \cite{Book} in a $c$-number part 
plus an operator  with vanishing expectation 
value 
is applied:  
$\hat{\psi}(\bbox{r},t)= \Phi(\bbox{r}) + \tilde{\psi}(\bbox{r},t)$.
$\Phi(\bbox{r})$ represents the condensate wave function and the operator 
$\tilde{\psi}(\bbox{r},t)$ the excitations of the condensate.

This ansatz is inserted in the Heisenberg equation of motion (in units of 
$\hbar=1$): 
\begin{equation}
\label{heq}
i\,\frac{\partial\hat{\psi}(\bbox{r},t)}{\partial t}=
\left( -\frac{\nabla^2}{2m}+U_{\text{trap}}(\bbox{r})-\mu\right)\hat{\psi}
(\bbox{r},t) +g\,\hat{\psi}^{\dagger}(\bbox{r},t) \hat{\psi}(\bbox{r},t)  
\hat{\psi}(\bbox{r},t)
\end{equation}
where $U_{\text{trap}}(\bbox{r})=
m(\omega^2_{\rho}\rho^2+\omega^2_z z^2)/2$ is the trap potential 
($m$ the atomic mass and $\omega_{\rho}$ and $\omega_z$ 
the radial and axial trap frequencies). 
The statistical  average over (\ref{heq}) and 
the replacement of  the cubic term
$\tilde{\psi}^{\dagger}(\bbox{r},t)
\tilde{\psi}(\bbox{r},t)\tilde{\psi}(\bbox{r},t)$ 
by the average in mean-field approximation
$2\left< \tilde{\psi}^{\dagger}(\bbox{r},t)
\tilde{\psi}(\bbox{r},t)\right>\tilde{\psi}(\bbox{r},t)$ 
neglecting the anomalous expectation value 
$\left< \tilde{\psi}(\bbox{r},t)
\tilde{\psi}(\bbox{r},t)\right>$ and its complex conjugate
leads to the  generalized Gross-Pitaevskii (GP) equation
in the HFB-Popov approximation
\begin{equation}
\label{gpe}
\left( -\frac{\nabla^2}{2m}+U_{\text{trap}}(\bbox{r})-\mu\right)\Phi(\bbox{r})
+g\left[ n_c(\bbox{r}) + 2 \tilde{n}(\bbox{r})\right]\Phi(\bbox{r}) = 0\, .
\end{equation}
Here we introduced the local density of the condensate $n_c(\bbox{r})=
|\Phi(\bbox{r})|^2$ and of the depletion 
$\tilde{n}(\bbox{r})=\left<\tilde{\psi}^{\dagger}(\bbox{r},t)\,
\tilde{\psi}(\bbox{r},t)\right>$.  
The subtraction of equation  (\ref{gpe}) from (\ref{heq}) 
and a treatment of the cubic terms within the
mean-field theory similar to the one of equation (\ref{gpe})
yields two  coupled equations of motion for  $\tilde{\psi}(\bbox{r},t)$
and its adjoint.
 
We only state  the final results 
after the insertion of  the
Bogoliubov transformation \cite{Fetter}: $\tilde{\psi}(\bbox{r},t)=\sum_j \,
[u_j(\bbox{r})\,\hat{\alpha}_j\, 
e^{-i E_j t}+ v_j^{*}(\bbox{r})\,\hat{\alpha}_j^{\dagger}\, e^{i E_j t}]$ 
leading to the 
generalized Bogoliubov equations
for the generalized 
eigenfunctions $u_j(\bbox{r})$,$v_j(\bbox{r})$ and their  corresponding 
eigenvalues $E_j$:
\begin{eqnarray}
\nonumber\hat{{\cal L}}\,u_j(\bbox{r}) + g\, n_c(\bbox{r})\, 
v_j(\bbox{r})&=&E_j \,u_j(\bbox{r}) \\
\label{beq}& & \\
\nonumber\hat{{\cal L}}\,v_j(\bbox{r}) + g \,n_c(\bbox{r})\, 
u_j(\bbox{r})&=&-E_j\, v_j(\bbox{r}) 
\end{eqnarray}
where $\hat{{\cal L}}\equiv -\frac{\nabla^2}{2m}+
U_{\text{trap}}(\bbox{r})-\mu+2\, g \,n(\bbox{r})$  
with the total density $n(\bbox{r})=n_c(\bbox{r})+\tilde{n}(\bbox{r})\,$.\\
The number density of particles $\tilde{n}(\bbox{r})$ outside the condensate 
is given 
in terms of the thermal number of quasi-particles 
$\left< \hat{\alpha}^{\dagger}_j\, \hat{\alpha}_j \right>
=\left( e^{E_j/k_{\text{B}}T} -1\right)^{-1} $ by 
\addtocounter{equation}{-1}
\begin{mathletters}
\label{ben}
\begin{equation}
\tilde{n}(\bbox{r})=\sum_j\left\{\left(|u_j(\bbox{r})|^2+|v_j(\bbox{r})|^2
\right)
\left< \hat{\alpha}^{\dagger}_j\, \hat{\alpha}_j\right>+|v_j(\bbox{r})|^2
\right\}\label{bena}
\end{equation}
\end{mathletters}
where $u_j(\bbox{r})$, $v_j(\bbox{r})$ are normalized by 
$\int \mbox{d}^3\bbox{r}\left(|u_j(\bbox{r})|^2-|v_j(\bbox{r})|^2\right)=1$.\\
The coupled equations can either be solved by 
direct expansion of the solution 
in the basis set of the free trap potential as in \cite{DoddEd} 
or by using the local density 
approximation as in \cite{Giorgini,Dalfovo}.
The local density approximation is the leading order of a semiclassical
approximation and amounts to setting
\begin{equation}
u_j(\bbox{r})\rightarrow u(\bbox{p}, \bbox{r})e^{i\phi(\bbox{r})} 
\qquad  v_j(\bbox{r})\rightarrow v(\bbox{p}, \bbox{r})e^{i\phi(\bbox{r})}
\qquad \sum_j\ldots \rightarrow 
\int\frac{\mbox{d}^3\bbox{p}}{\left(2\pi\right)^3}\ldots\, .
\end{equation}
where $\phi$ is defined by
$\nabla\phi\equiv \bbox{p}$ and $u(\bbox{p},\bbox{r})$, 
$v(\bbox{p},\bbox{r})$ are normalized by \\
$|u(\bbox{p},\bbox{r})|^2-|v(\bbox{p},\bbox{r})|^2=1$.\\
In the semiclassical limit the functions 
$ u(\bbox{p},\bbox{r})$ and $v(\bbox{p},\bbox{r})$ are slowly varying on 
the scale of 
the harmonic oscillator length $\bar{d}=(1/m\bar{\omega})^{1/2}$ 
with 
$\bar{\omega}=(\omega_{\rho}^2\,\omega_z)^{1/3}$, and 
the derivatives of $u$ and $v$ as well as the second derivatives
of $\phi$ are negligible. 
The equations in (\ref{beq}) are then reduced to  the algebraic form

\begin{eqnarray}
\nonumber
\left(\frac{p^2}{2m} + U_{\text{trap}}(\bbox{r}) -\mu +2 
g n(\bbox{r})\right) u(\bbox{p},\bbox{r}) 
+ g n_c(\bbox{r}) v(\bbox{p},\bbox{r}) &=&\epsilon (\bbox{p},\bbox{r}) 
u(\bbox{p},\bbox{r})\\
\label{ldbeq} \\
\nonumber
\left(\frac{p^2}{2m} + U_{\text{trap}}(\bbox{r}) -\mu +2 
g n(\bbox{r})\right) v(\bbox{p},\bbox{r}) 
+ g n_c(\bbox{r}) u(\bbox{p},\bbox{r}) &=&-\epsilon (\bbox{p},\bbox{r}) 
v(\bbox{p},\bbox{r})\,.
\end{eqnarray}

The solution is straight-forward and we are led to the excitation spectrum  
$\epsilon (\bbox{p},\bbox{r})=(\epsilon_{HF}^2(\bbox{p},\bbox{r})-g^2 
n_c^2(\bbox{r}))^{1/2}$ with the Hartree-Fock energy $\epsilon_{HF}
(\bbox{p},\bbox{r})=\frac{\bbox{p}^2}{2m}+U_{\text{trap}}(\bbox{r})-
\mu+2gn(\bbox{r})$.
The non-condensate density is then obtained by a simple integration over 
the momenta:
\begin{eqnarray}
\label{ntlda}
\tilde{n}(\bbox{r})&=\displaystyle
\int\frac{\mbox{d}^3\bbox{p}}{(2\pi)^3}&
\left[\frac{\epsilon_{\text{HF}}(\bbox{p},
\bbox{r})}{\epsilon(\bbox{p},\bbox{r})}
\left(\frac{1}{\exp(\epsilon(\bbox{p},\bbox{r})/
k_B T)-1}+\frac{1}{2}\right)-\frac{1}{2}
\right] 
\Theta(\epsilon^2_{\text{HF}}(\bbox{p},\bbox{r})-g^2n_c^2(\bbox{r})
)\,\,\,.
\end{eqnarray}
Let us first consider the case where the local density approximation is used 
for the whole excitation spectrum, not only the high-lying part.
Then it is necessary for consistency to treat also 
the condensate in the corresponding approximation, which is the 
finite-temperature Thomas-Fermi approximation.
It applies to sufficiently large condensates.
In this case the kinetic energy term in eq.(\ref{gpe}) is neglected and one
obtains with $n(\bbox{r})=\tilde{n}(\bbox{r})+n_c(\bbox{r})$  
\begin{equation}
\label{conddens}
n_c(\bbox{r})=g^{-1}\left(\mu-U_{\text{trap}}(\bbox{r}) 
-2 g \tilde{n}(\bbox{r})
\right)\Theta\left(\mu-U_{\text{trap}}(\bbox{r})
-2 g \tilde{n}(\bbox{r})\right)
\end{equation}
and hence $\epsilon(\bbox{p},\bbox{r})$ reduces to 
\begin{equation}
\label{ldae}
\epsilon(\bbox{p},\bbox{r})=
\left\{
\begin{array}{ccc}
\left(\left(\frac{\bbox{p}^2}{2m}+g n_c(\bbox{r})\right)^2-
g^2n_c^2(\bbox{r})\right)^{1/2} &\mbox{if} & \mu>U_{\text{trap}}(\bbox{r})
+ 2 g \tilde{n}(\bbox{r}) \\
\frac{\bbox{p}^2}{2m}+U_{\text{trap}}(\bbox{r})-\mu+2 g n(\bbox{r})
&\mbox{if} & \mu < U_{\text{trap}}(\bbox{r})
+ 2 g \tilde{n}(\bbox{r}) 
\end{array}
\right.
\end{equation}
Inside the condensate the excitation energy is then gapless like in the 
spatially homogeneous case.
Moreover the integrand in (\ref{ntlda}) for $\bbox{r}$ inside the condensate
and $p\rightarrow 0$ becomes $k_B T \,m/p^2$ 
and  therefore the integral (\ref{ntlda}) converges 
for small $p$ (besides, of course, converging also for large $p$).

On the other hand, even if the Thomas-Fermi approximation for the
condensate is not applicable because the condensate is to small 
one may still apply the semiclassical local density approximation 
for sufficiently high lying states. But for the low-lying states the 
local density approximation is then inconsistent, as can e.g. be seen 
from the fact that $\epsilon(\bbox{p},\bbox{r})$ inside the condensate 
now has a gap for $p\rightarrow 0$ of size $E_{\text{gap}}(\bbox{r})=
\left(
\frac{\nabla^2\Phi(\bbox{r})}{2m\Phi(\bbox{r})}
\left(\frac{\nabla^2\Phi(\bbox{r})}{2m\Phi(\bbox{r})}+2 g n_c(\bbox{r})\right)
\right)^{1/2}$ in space points $\bbox{r}$ 
where $\nabla^2\Phi(\bbox{r})>0$.
Therefore, one has to introduce an energy cut-off $\epsilon_c$ in this case, 
below which
the excitations are treated by solving eqs.(\ref{beq}) exactly, 
expanding
in a suitable basis, and above which the local density approximation can 
still be employed.
Then eq.(\ref{ntlda}) is replaced by 
\begin{equation}
\tilde{n}(\bbox{r})=\sum_j\tilde{n}_j(\bbox{r})\,\Theta(\epsilon_c-E_j)
+\int_{\epsilon_c}^\infty\mbox{d}\epsilon\,\tilde{n}(\epsilon,\bbox{r})
\label{ntldana}
\end{equation}
where
\begin{equation}
\tilde{n}_j(\bbox{r})=\frac{|u_j(\bbox{r})|^2+|v_j(\bbox{r})|^2}{
e^{E_j/k_{\text{B}}T}-1} +|v_j(\bbox{r})|^2\label{ntldanb}
\end{equation}
and
\begin{eqnarray}
\nonumber
\tilde{n}(\epsilon,\bbox{r})&=&\frac{m^{3/2}}{\sqrt{2}\pi^2}\left\{
\frac{1}{e^{\epsilon/k_{\text{B}}T}-1}+\frac{1}{2}-\frac{\epsilon}{2\sqrt{
\epsilon^2+g^2n_c^2(\bbox{r})}}\right\}\\
& &\times
\sqrt{\sqrt{\epsilon^2+g^2n_c^2(\bbox{r})}-U_{\text{trap}}(\bbox{r})
+\mu-2gn(\bbox{r})}\label{ntldanc}
\end{eqnarray}

It is our goal here to implement this hybrid procedure.
The choice of the energy cut-off depends on the purpose of the calculation.
If the aim is the calculation of thermodynamic properties requiring only
the density of states it is sufficient to take the energy cut-off low
at an energy of a few $\bar{\omega}$,  
and to restrict the discrete part of the sum over states to 
only very few discrete modes.
If on the other hand the goal is to calculate the mode spectrum 
of the low-lying modes accurately up to a given energy, then this  energy 
determines the cut-off. However, the employed basis set must then still
be chosen considerably larger (in practice about twice as large) than
the number of discrete modes to be calculated, in order to make sure that no
discrete mode below the cut-off is missed.
In this way it is ensured that 
eq. (\ref{ntldana}) really gives an 
accurate value of the particle density outside the condensate.
The advantage of our hybrid approximation over the hitherto used simple
cut-off of the employed basis set is that the truncation error in 
$\tilde{n}(\bbox{r})$ is avoided, which allows us the use
of a smaller basis set and therefore a more economical computation.

\section{COMPUTATION}
The self-consistent procedure we use combines and extends previously used
methods of ref.
\cite{Edwards,Giorgini,HutchZarGriffin}.
For a given number of atoms, temperature and anisotropy of the trap 
we start e.g. with
$\tilde{n}=0$ or a previously determined better estimate for $\tilde{n}$,
solve eq.(\ref{gpe}) for $\Phi(\bbox{r})$, $n_c(\bbox{r})=|\Phi(\bbox{r})|^2$
and $\mu$ from the condition that $N=\int\mbox{d}^3\bbox{r}\left(n_c(\bbox{r})
+\tilde{n}(\bbox{r})\right)$, solve then eq.(\ref{beq}) for all states 
up to the chosen 
energy cut-off to find $u_j(\bbox{r})$, $v_j(\bbox{r})$ and $E_j$,
and determine finally an improved value for $\tilde{n}(\bbox{r})$ from 
eqs.(\ref{ntldana}), (\ref{ntldanb}).

To calculate the second term in eq.(\ref{ntldana}) it is necessary to 
introduce a finite region in space, where we perform the energy integral at 
certain points. Due to the axial symmetry of our problem this finite region is 
a rectangle in the $r$ and $z$ plane. This rectangle is chosen in such 
a way that
$\tilde{n}(\bbox{r})$ and $n_c(\bbox{r})$ fall fast enough to zero approaching
the edges.
In this rectangle we introduce a fine grid and the energy integrals are 
performed at every grid point numerically. Since both, $n_c(\bbox{r})$ and
$\tilde{n}(\bbox{r})$ are smoothly varying functions the grid methods are 
suitable from the numerical point of view. We use the same grid in solving the
Gross-Pitaevskii eq.(\ref{gpe}).
To reach fast convergence for the solution of (\ref{gpe}) we apply the
Newton-Raphson algorithm in combination 
with multigrid methods as discussed in ref. 
\cite{Hackbusch}.

However, the grid methods are not appropriate in determining the solutions
of eq.(\ref{beq}) especially for the high-lying states.
For this reason we solve the Bogoliubov-equations in the basis 
of the eigenfunctions of the free trap.
In this basis we only need to calculate numerically the matrix elements 
of  $g\,n_c(\bbox{r})$, $g\,\tilde{n}(\bbox{r})$ in harmonic oscillator 
eigenstates. These matrix elements are decaying fast 
enough, because these quantities are smooth and localized around the origin.
For our purposes the best method to solve (\ref{beq}) in our basis 
is that of Hutchinson et al. \cite{HutchZarGriffin}.
Then we do not have to introduce a basis for $u_j$ and $v_j$ separately, 
a single basis set is enough but two subsequent diagonalizations 
must be performed. For further details see \cite{HutchZarGriffin}. 

\section{RESULTS}

We shall now discuss some results obtained by the procedure described in the
previous section.
In the calculations we present here we choose $N=2000$ Rubidium atoms 
and $T=0.5\, T_c$
where $T_c$ is the critical temperature of an ideal Bose-gas with the same 
parameters. The anisotropy of the trap is described by  the parameter
\begin{equation}
\label{betadef}
\beta=2\frac{\omega_z^2-\omega_{\rho}^2}{2\omega_{\rho}^2+\omega_z^2}\,,
\qquad\omega_{\rho}=\omega_{av}\sqrt{1-\frac{\beta}{2}}\,,
\qquad\omega_z=\omega_{av}\sqrt{1+\beta}
\end{equation}
where $\omega_{av}$ is fixed to $\omega_{av}=137\times 2\pi$ Hz.
As a check on our code we reproduced and confirm the zero-temperature results
for the low-lying excitation spectrum as a function of the 
anisotropy parameter presented in fig.1 of ref.\cite{HutchZar}.
A further check on our code, now at finite temperature,
is the agreement of our results for the low-lying excitation spectrum as 
a function of temperature for the special anisotropy of the JILA-TOP trap 
with the results in \cite{DoddEd}.

Let us now turn to our own results. In fig.~\ref{fdnde} 
we fix $\beta$ at $\beta=1.4$
and plot the spectral distribution $g(\epsilon)$ of the particles
in the trap over energy $\epsilon$ in units of $\omega_{\rho}$.
Thus $\text{d}N=g(\epsilon)\text{d}\epsilon$ is the number of atoms
with energy $\epsilon$ in the interval $\text{d}\epsilon$.
In fig.~\ref{fdnde} we compare the spectral distribution 
\begin{mathletters}
\label{gs}
\begin{equation}
g_{\text{ld}}(\epsilon)=
\int \mbox{d}^3\bbox{r}\,\tilde{n}(\epsilon,\bbox{r})\,\Theta\left(\epsilon
-U_{\text{trap}}(\bbox{r})+\mu-2gn(\bbox{r})\right)\label{gsa}
\end{equation}
obtained in the
local density approximation with $\tilde{n}(\epsilon,\bbox{r})$ from
eq.(\ref{ntldanc}) (full line), with the spectral distribution 
\begin{equation}
g_{\text{d}}(\epsilon)\equiv\displaystyle
\sum_j\int\mbox{d}^3\bbox{r}\,\tilde{n}_j(\bbox{r})
\frac{1}{\sqrt{\pi\omega_{\rho}^2}}
\exp\left(-(\frac{\epsilon-E_j}{\omega_\rho})^2
\right)\label{gsb}
\end{equation}
\end{mathletters}
 obtained from the low-lying discrete modes by folding 
$\tilde{n}_j(\bbox{r})$
(\ref{ntldanb}) with a Gaussian of variance 
$\omega_{\rho}/\sqrt{2}$. 
Some smoothing on the quantum scale is necessary
for a meaningfull comparison. The energy cut-off was chosen at 
$\epsilon_c=30\,\omega_{\rho}$, 
which is the reason for the fall-off of $g_d$
at around this energy. It can be seen that the local density approximation 
is rather good for energies somewhat above the chemical potential and higher.
For energies at $\mu$ or lower the local density approximation increasingly
fails and the spectral distribution from the discrete modes must be used.
The oscillations by which $g_{\text{d}}(\epsilon)$ differs from 
$g_{\text{ld}}(\epsilon)$ may be understood semiclassically \cite{Gutzwiller}
by analysing the short periodic orbits of the classical dynamics generated
by the Hamiltonian $\epsilon({\bbox p}, {\bbox r})$ \cite{Fliesser2}.

Next we discuss results for the excitation spectrum at temperature
$T=0.5 \,T_c$ as a function of the anisotropy parameter $\beta$. We use the 
same parameter values as chosen in ref.\cite{DoddEd} where the 
excitation spectrum for the special value of $\beta=1.4$ of the JILA-TOP trap 
was presented. This case is therefore contained in our results.
In fig.~\ref{fmu} we show the behavior of the chemical potential 
$\mu$ for
$N=2000$ atoms as a function of $\beta$. In the vicinity of the limiting cases 
$\beta=-1,2$ the chemical potential changes rapidly and is expected to tend 
towards the groundstate energies of the free trap.

The parity quantum number and the azimuthal quantum number $m$ are 
good quantum numbers for axially symmetric harmonic traps.
On figs.~\ref{fevev}-\ref{fodev} we show the elementary excitations 
as a function of $\beta$ for different parity classes and $m$. 
For clarity only 
the excitation spectra with the quantum numbers $m=0,1,2,3$ are presented.
On Figures~\ref{fevod} and \ref{fodev} the lowest curves 
show  the behavior of the center of mass modes, for odd parity with $m=0$ and 
for even parity with  $m=1$. 
They represent the collective oscillations of the atoms 
with the harmonic frequencies of the free trap $\omega_z$
and $\omega_{\rho}$.
By Kohn's theorem \cite{Kohn} they must occur in all finite temperature 
spectra and would be straight lines in our pictures.
From our numerical data we 
see that Kohn's theorem is not exactly maintained in the HFB-Popov
approximation,
but the discrepancy for all anisotropies is so small, that
it is practically negligible at $T=0.5\,T_c$.
In the excitation spectra (figs.~\ref{fevev}-\ref{fodev})
modes with different axial quantum numbers $m$ cross freely
as one would expect if (\ref{beq}) is considered as a linear set of 
eigenvalue equations.
However, due to the coupling of these equations via $\tilde{n}(\bbox{r})$,
which depends on the $|u_j(\bbox{r})|^2$ and the $|v_j(\bbox{r})|^2$,
the linearity is actually broken for finite temperature.
But the nonlinearity enters only via a global coupling.
Therefore, in discussing the crossings of levels we may consider 
$n(\bbox{r})$ and $n_c(\bbox{r})$ as  given by their final self-consistent
values in which case (\ref{beq}) does become linear explaining the 
free level-crossings for different quantum numbers $m$.
Some avoided crossings between levels with the same $m$ can also be seen 
in these figures.

A remarkable feature of the above spectra is that the levels tend 
to a few common eigenvalues in the limits $\beta\rightarrow2$ 
or $\beta\rightarrow-1$.
In the first case ($\beta=2$) $\omega_{\rho}$ goes to zero and the 
limiting energy eigenvalues  
do not depend on the azimuthal quantum number $m$ but
still on the parity. In the opposite case ($\beta=-1$) 
they now depend on $m$ but not on the parity.
However, in the limit $\beta=2$ the spectra  
agree with that of the 
free harmonic oscillator
with $\omega_z=0$. This fact can be easily understood because both, the
condensate and the thermal density vanish due to the repulsion 
between particles and the vanishing confinement in radial direction.
The same argument can be applied for $\beta=-1$ with vanishing 
confinement in the $z$-direction.
However, in the close neighborhood of $\beta=-1$ some quantities
(see for example the behavior of the chemical potential in fig.~\ref{fmu}) 
change too rapidly and it is extremely difficult to
get reliable data for the spectra in that region. 
Note that the limiting situations are quite different from those
studied by Stringari \cite{Str2} (see also ref.~\cite{Fliesser})
under the conditions that the hydrodynamic approach is applicable, 
though some features are common.

The low-lying excitation spectrum as a function of the anisotropy 
parameter for zero temperature has been presented by 
{\it Hutchinson et al} \cite{HutchZar}.
The results are remarkably similar to ours considering the 
widely different physical conditions under which they are obtained.
The similarity of the results of finite temperature calculations in 
the HFB-Popov approximation to zero-temperature results was first 
noted by {\it Dodd et al} \cite{DoddEd}.
For anisotropy $\beta=1.4$ they compared their results with an effective
zero-temperature calculation  (EZC), where $N=N_c$, using as effective
particle number $N$ the number of particles 
in the actual condensate at temperature $T$.
The close agreement they found led to the conclusion that in practice
the excitation spectra are determined mainly by the atoms in the 
condensate, so that the thermal density $\tilde{n}(\bbox{r})$ can be 
neglected in the GP and the Bogoliubov equations (\ref{gpe}) and
(\ref{beq}).
Furthermore, at least for the high-lying excitations 
\cite{HutchZarGriffin,Bergemann,Houbiers} 
$v_j({\bf r})$ can be set to zero in (\ref{beq}) 
\cite{HutchZarGriffin,Bergemann} and the equations  (\ref{beq}) reduce 
to an ordinary Schr\"odinger equation.
We can now use our result for arbitrary values of $\beta$ and looking 
also at higher lying excitation energies to 
provide a much more sensitive test of the accuracy of the EZC model.

In fig.~\ref{fezc}
we present the low-lying EZC mode spectra for the same 
case as in fig.~\ref{fevev}. In the EZC approximation the Kohn-theorem is 
satisfied exactly, as there is no thermal cloud in the effective
zero-temperature system. 
The excellent agreement with the HFB-Popov results, even in fine details,
over the whole interval of anisotropies confirms the remarkable usefulness
of the EZC approximation for the calculation of the low-lying
excitation spectrum. 

However, considering a broader range of energies one can find
differences between the predictions of the HFB-Popov and the 
EZC-calculations. In order to make a quantitative comparison we fix the 
anisotropy parameter $\beta$ in fig.~\ref{fcom}
to the value of the JILA trap $\beta=1.4$. Here we plot
the energy-differences between the HFBP-
and the EZC-calculations $\Delta \epsilon_1=\epsilon_{\text{HFBP}}
-\epsilon_{\text{EZC}}$ and between the HFBP- and the 
Hartree-Fock-calculations $\Delta \epsilon_2=\epsilon_{\text{HFBP}}
-\epsilon_{\text{HF}}$ as a function of the HFBP-excitation energies.
At very low energies $\Delta \epsilon_1$ is very small 
confirming again that in the low-lying spectra 
EZC levels agree with the HFB-Popov levels.
With increasing energy the EZC reveals 
an increasing shift and $\Delta\epsilon_1$ shows larger fluctuations.
The difference between the
eigenvalues is less than $0.2\omega_{av}$. It is not too big, but
shows clearly the magnitude of the difference between the two calculations.
For energies above twice the chemical potential 
the Hartree-Fock model ($v_i$ neglected, but including the thermal
density $\tilde{n}$) turns out to be a 
better approximation for the HFB-Popov equations.
The shift for large energies can be understood in 
the following way: There is a shift in the chemical potential 
$\Delta \mu= \mu_{\text{HFB}}-\mu_{\text{EZC}}=0.19$ because of the 
neglection of the
almost constant thermal density in the condensate region in the EZC.
Considering only small energies the eigenmodes are 
localized inside the  condensate where  the term $2g\tilde{n}$
and the chemical potential shift 
cancel each other.
Therefore the energy difference $\Delta \epsilon_1$ is small 
for low energies. But for states 
localized outside,  where even $\tilde{n}$ is negligible the
difference $\Delta \mu$ is the main reason for the shift of the levels
in the EZC calculations.  The Fig.~\ref{fcom} is in accord with the
expectation that the assymptotic limit of the energy shift is 
$\Delta \mu$.

\section{CONCLUSIONS}
We have investigated the Hartree-Fock-Bogoliubov-Popov equations
to obtain  the finite temperature excitation spectrum 
of trapped condensed Bose gases. Our computational method combined the use
of the local density approximation 
above a certain cut-off and a solution of the discrete Bogoliubov equations 
below the energy cut-off.
We have explored in a systematic way the excitation frequencies within a large
range of parameters for the anisotropy of the trap potential.
We gave numerical evidence for degeneracies occuring at extreme anisotropic
situations $\beta=2,-1$ and argued that these two spectra can be explained
by the non-interacting model.
Further we compared the spectra of the HFB-Popov with the EZC calculations
and showed that for large energies the EZC spectra are shifted by a small
amount
which is mainly due to the shift in the chemical potentials.

\acknowledgments
This work has been supported by the project of the Hungarian 
Academy of Sciences and the Deutsche Forschungsgemeinschaft under Grant No. 95.
R.~G. and J.~R. wish to acknow\-ledge support by the Deutsche 
Forschungsgemeinschaft through the Sonderforschungsbereich 237 ``Unordnung und 
gro{\ss}e Fluktuationen''. Two of us (A.~Cs. and P.~Sz.) would like 
to acknowledge
support by the Hungarian National Scientific Research Foundation under Grant 
Nos. OTKA T025866, T017493 and F020094 and by the Ministry of Education
of Hungary under grant No. FKFP0159/1997.

\bibliographystyle{plain}

\begin{figure}
\caption{The spectral distribution 
$g_{\text{ld}}(\epsilon)$ (\ref{gsa})
in the local
density approximation is compared with the spectral distribution $g_{\text{d}}
(\epsilon)$ (\ref{gsb}) obtained by the expansion in a basis set
for $\beta=1.4$, $N=2000$, $T=0.5\, T_c$.} 
\label{fdnde}
\end{figure}

\begin{figure}
\caption{$\mu$ in units of $\omega_{av}$ 
as a function of the anisotropy parameter $\beta$ for $N=2000$, 
$T=0.5 T_{\text{c}}$.}
\label{fmu}
\end{figure}

\begin{figure}
\caption{The lowest HFB-Popov excitations $\omega^2$  
in units of $\omega_{av}^2$
as a function of $\beta$ 
for the even-parity modes
with $m=0$ (open circles), 
$m=2$ (black stars). Other parameters are
$N=2000$, $T=0.5\,T_c$.} 
\label{fevev}
\end{figure}

\begin{figure}
\caption{The same as in fig.~{\protect\ref{fevev}} for the even-parity
modes $m=1$, $m=3$.}
\label{fevod}
\end{figure}

\begin{figure}
\caption{The same as in fig.~{\protect\ref{fevev}} for the odd-parity
modes $m=0$, $m=2$.}
\label{fodev}
\end{figure}

\begin{figure}
\caption{The lowest {\it equivalent zero condensate}  excitations $\omega^2$ 
corresponding to fig.~{\protect\ref{fevev}}  
in units of $\omega_{av}^2$
as a function of $\beta$ 
for the even-parity modes
with $m=0$ (open circles), 
$m=2$ (black stars).}
\label{fezc}
\end{figure}

\begin{figure}
\caption{The energy differences $\Delta\epsilon_1
=\epsilon_{\text{HFBP}}
-\epsilon_{\text{EZC}}$ (open circles) and  
$\Delta\epsilon_2=\epsilon_{\text{HFBP}}
-\epsilon_{\text{HF}}$ (crosses) 
as a function of the HFBP-energies.} 
\label{fcom}
\end{figure}

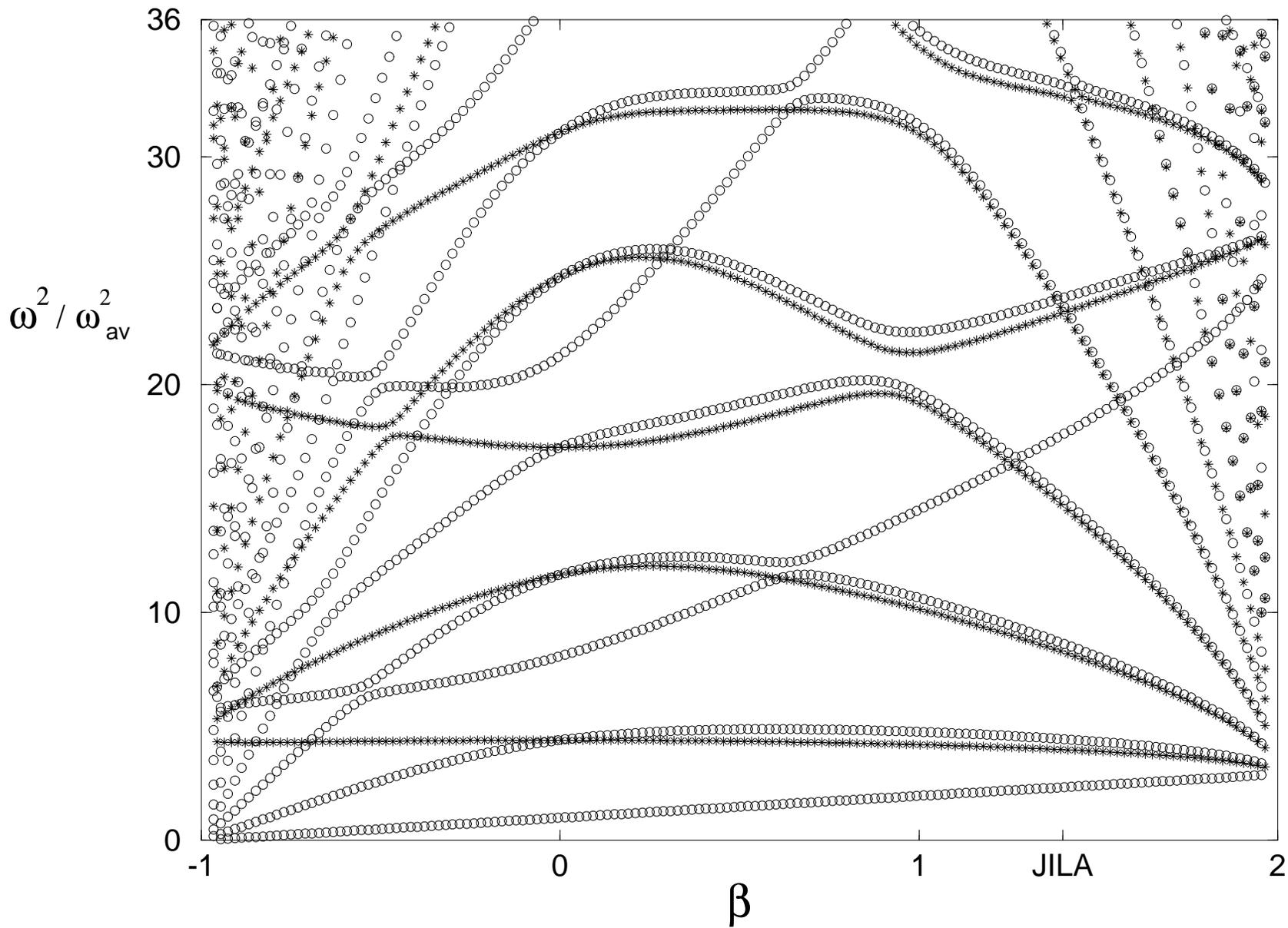
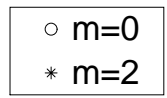
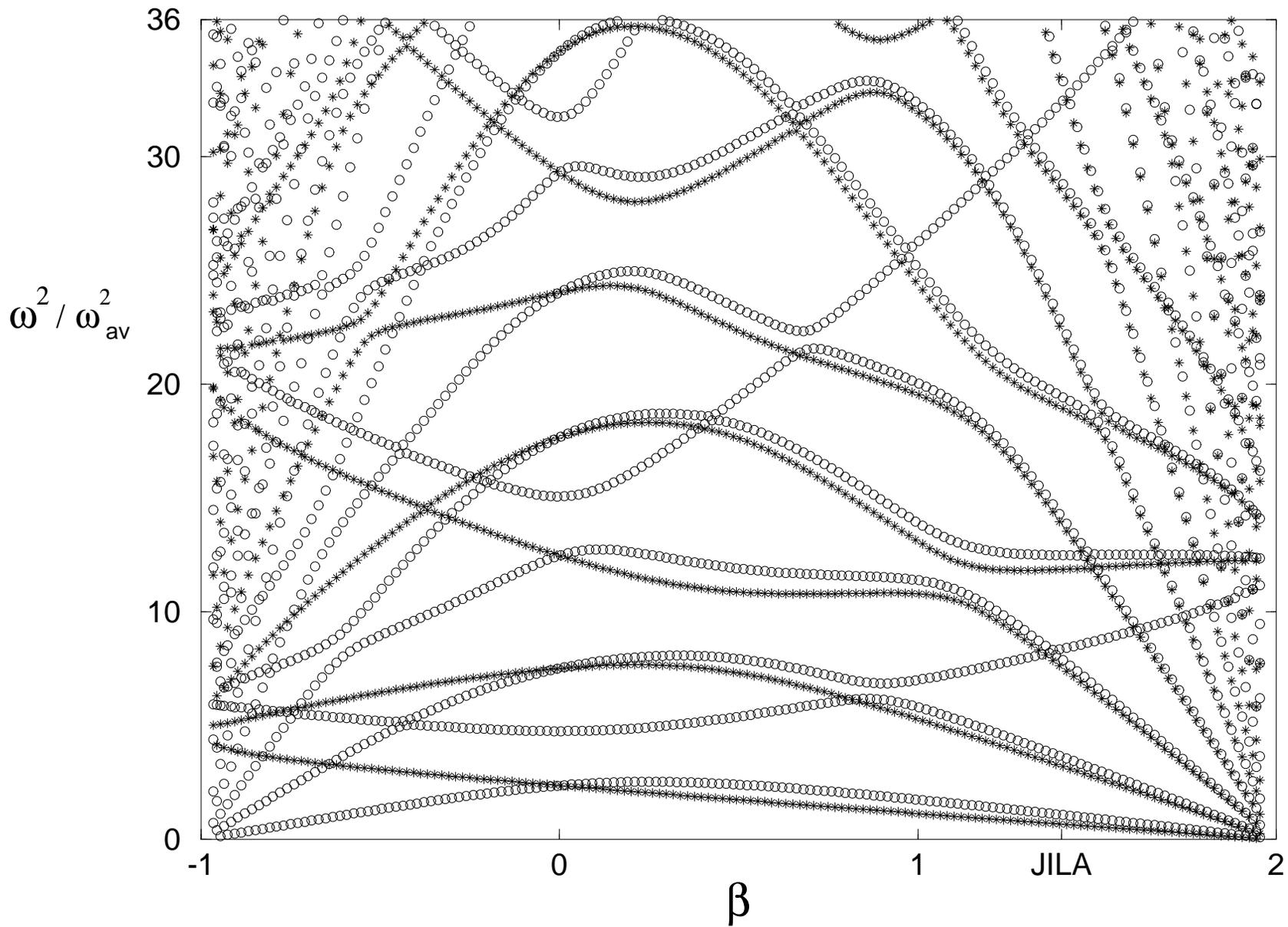
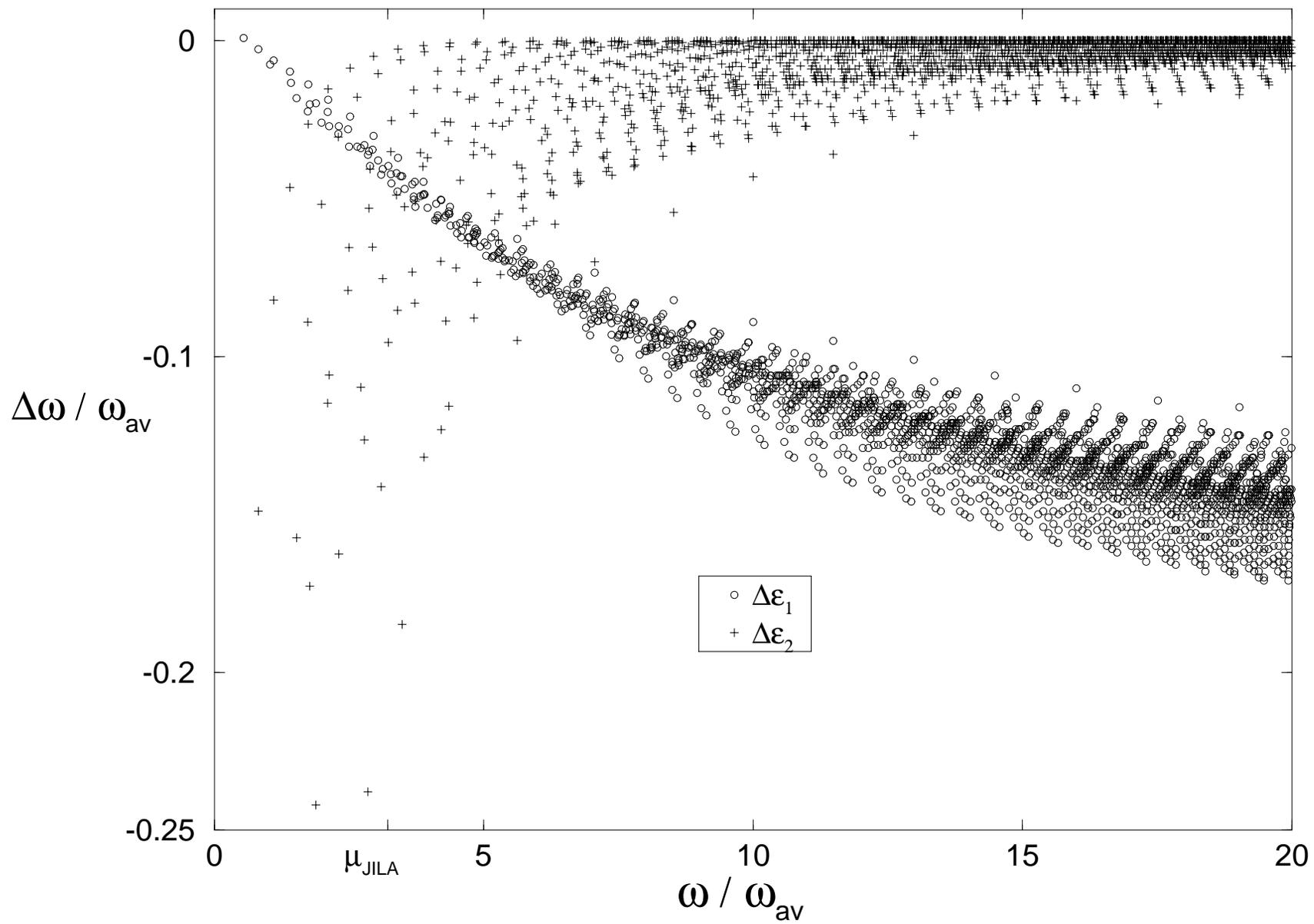
\end{document}